# Magnetic Hyperthermia with $Fe_3O_4$ nanoparticles: the Influence of Particle Size on Energy Absorption.

Gerardo F. Goya[1], Enio Lima, Jr.[2], Amanda D. Arelaro[2], Teobaldo Torres[1], Hercilio R. Rechenberg[2], Liane Rossi[3], Clara Marquina[4], and M. Ricardo Ibarra[1,4]

[1]Instituto de Nanociencia de Aragón (INA), University of Zaragoza, Spain.
[2]Instituto de Física - Universidade de São Paulo, São Paulo, Brazil.
[3]Instituto de Química – Universidade de São Paulo, São Paulo, Brazil.
[4]Instituto de Ciencia de Materiales de Aragón-CSIC, University of Zaragoza, Spain.

**We have studied the magnetic and power absorption properties of a series of magnetic nanoparticles (MNPs) of $Fe_3O_4$ with average sizes <d> ranging from 3 to 26 nm. Heating experiments as a function of particle size revealed a strong increase in the specific power absorption (SPA) values for particles with <d> = 25-30 nm. On the other side saturation magnetization $M_S$ values of these MNPs remain essentially constant for particles with <d> above 10 nm, suggesting that the absorption mechanism is not determined by $M_S$. The largest SPA value obtained was 130 W/g, corresponding to a bimodal particle distribution with average size values of 17 and 26 nm.**

## I. Introduction

THE CAPABILITY of magnetic nanoparticles (MNPs) to act as effective heating agents for Magnetic Hyperthermia (MHT) has been demonstrated many years ago.[1] However, the mechanisms and true limits for delivering the absorbed power to small intracellular structures are still open questions.[2] The therapeutic approach of MHT consist in remotely raising the temperature of a target tissue through alternating magnetic fields acting on magnetic nanoparticles previously loaded in that tissue. The safeness of MHT as compared with microwave or conduction-based hyperthermia is related to the frequency region ($f = 10^2 - 10^3$ kHz) of the electromagnetic radiation used by MHT, where the heating effects on living tissues are negligible.

The origin of the energy dissipation when a magnetic colloid is placed on an ac magnetic field has been subject of extensive work, but experimental difficulties for generating ac magnetic fields of tens of mT at different frequencies have hindered accurate models to be developed. Moreover, most of the experimental and theoretical works reported so far have dealt with single-domain particles, and thus the discussion of absorption models has been restricted to Brownian and Néel relaxation mechanisms. The prevailing picture today is that power absorption takes place mainly through Néel relaxation of the magnetic moments, since energy losses from mechanical rotation of the particles, acting against viscous forces of the liquid medium (Brownian losses) cannot contribute at those frequencies used for MHT ($10^2 - 10^3$ kHz).

However, for iron-oxide particles with sizes larger than c.a. 25-30 nm, a transition from single- to multi-domain structure takes place. For such MNPs the existence of a multidomain structure within a ferromagnetic nanoparticle imply that magnetic losses through domain wall (DW) displacements can contribute to power absorption, in the same way that bulk ferromagnetic materials do. Since initial susceptibility of a ferromagnetic material is much larger than for a superparamagnetic particle, DW contribution could be larger than Néel relaxation if large field amplitudes are used, allowing nanoparticles to act as more efficient nanoheaters. In this work, we undertake the systematic study of $Fe_3O_4$ nanoparticles with controlled size from 3 to 25 nm, in order to determine the relevant parameters involved in the absorption power of MNPs as the average particle size approaches the multi-domain critical value.

## II. Experimental Procedure

### A. Sample Preparation

Samples composed of MNPs with <d> < 10 nm were synthesized by single-step high-temperature decomposition of the precursor $Fe(acac)_3$ in the presence of a long-chain alcohol as reported by Sun et al. [3]. The protocol was modified to tailor the final particle sizes by changing the molar ratio between the metallic precursor and the surfactant [4]. The solvent used was phenyl ether (boiling point ~543 K) and the synthesis lasted for 120 minutes in argon flux (~ 0.5 L/min.). The resulting nanoparticles were very stable against agglomeration because of the surfactant molecules attached to the surface of the magnetic cores. Three samples with <d> = 3.2, 7.3 and 9.5 nm were synthesized by the single-step procedure, labeled AP01, AP02 and AP03 respectively. Three more samples were grown using previously synthesized seeds (~ 10 nm) and the same protocol, in order to further increase the final particle size. These samples were labeled as GEYY, where YY indicates the number of times the re-crystallization took place (running from 01 to 03). The sequence goes like this: particles from sample AP03 were used as seeds for GE01 synthesis, GE01 as seeds of the GE02, and finally GE02 as seeds for GE03. We used 50 mg of seeds to 2 mmol of precursor, except for sample GE03 where we used 20 mg of seeds. Since the reaction time at high-temperature is added for each sample, the total crystallization time increases from 120 minutes (10 nm seeds) to 480 minutes for GE03.

*B. Particle Characterization*

High-resolution TEM (HRTEM) images were made in a Philips CM200 (200 kV) electron microscope, after dropping the colloidal solution onto a carbon-coated copper grid. X-ray diffraction (XRD) patterns were collected in θ-2θ geometry with a Philips W1700 diffractometer, using Cu-Kα radiation (λ = 0.15406 nm). Magnetization and ac susceptibility measurements were performed as a function of temperature from 2 to 300 K and applied fields up to 5 Tesla with a commercial SQUID magnetometer. Zero-field-cooling (ZFC) and field-cooling (FC) protocols were used when necessary. For ac susceptibility the frequencies were set from 0.1 to 10 kHz and field amplitude of 2 G. Specific Power Absorption (SPA) data was taken using a commercial ac applicator (model DM100 by nB nanoscale Biomagnetics) working at $f$ = 220-260 kHz and field amplitudes from 0 to 20 mT, and equipped with an adiabatic sample space (~ 0.5 ml) for measurements in liquid phase. Temperature data was taken using a fiber optic temperature probe (Reflex[TM], Neoptix) immune to rf environments.

III. EXPERIMENTAL RESULTS

X-ray diffraction of all particles could be indexed using the cubic $Fe_3O_4$ phase (space group Fd3m), as shown in Figure 1. An estimation of the average crystallite size was obtained from the most intense peaks by using the Scherrer equation, giving the values displayed in Table I. Accordingly, TEM images showed that the average particle size increased with increasing molar precursor/surfactant ratio. For the AP01 to AP03 samples (single-step synthesis) the morphology observed was mainly rounded (fig. 1), with a narrow size distribution. For samples GEYY, the morphology inferred from TEM was mainly cubic, and the size dispersion increased for each successive synthesis step. Sample GE03 displayed a bimodal distribution of particles, one centered at d ≈ 17 nm and a second with larger particles with average size of d ≈ 26 nm. The smaller particles in this sample correspond to the size of the seeds used as precursors of the new synthesis, indicating that the bimodal distribution is originated from original seeds that stay without reacting. One possible cause of this behavior could be a too low precursor concentration.

The presence of two distinct populations of MNPs in sample GE03 was further checked through ac susceptibility measurements (fig. 2), which showed two blocking temperatures located at ≈ 30 and ≈230 K for particles with 17 and 26 nm, respectively. The Arrhenus plot of the $1/T_B(f)$ data, used to extract the values of the effective anisotropy constant $K_{eff}$, yielded the same value for both blocking

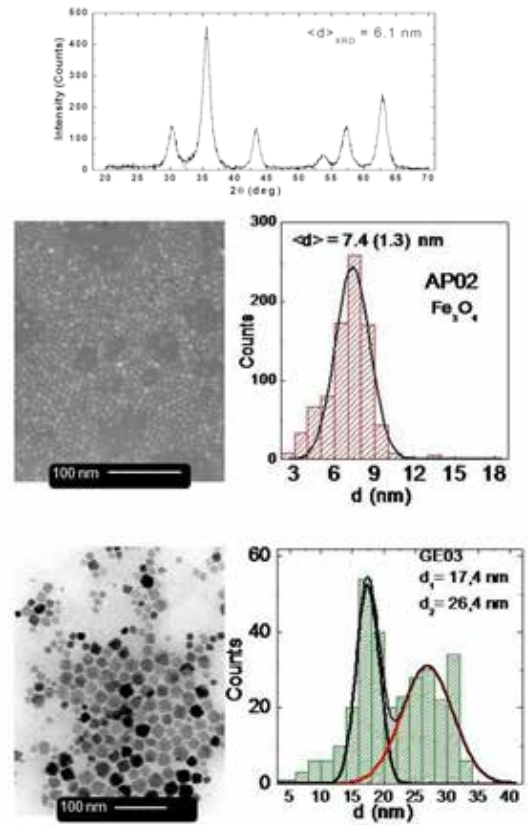

Fig. 1. Top panel: Typical X-ray diffraction pattern obtained for the sample series (AP02 in the figure above), showing the main peaks indexed as $Fe_3O_4$ phase. The peak at 2θ = 35.6º was fitted (red line) obtain the average grain size $<d>_{XRD}$ through the Scherrer equation in all samples. Middle panel: TEM image of same sample (left), and the corresponding histogram (right) fitted with a log-normal distribution to estimate the mean particle size $<d>_{TEM}$. Bottom panel: Tem and histogram for GE03 bimodal sample.

processes (Table I).

The saturation magnetization $M_S$ values at room temperature of all samples are lower than the expected 92-98 emu/g of bulk magnetite, but increase as the average particle size increases. This fact suggests that surface disorder is the origin of the diminished magnetization [5]. Figure 3 shows that the evolution of $M_S$ with $<d>$ displays a jump within the 7-10 nm range, and above these values it remains nearly constant. On the other hand, the SPA values vs. $<d>$ remain essentially zero small up to 15 nm, and for the sample with bimodal distribution of 17 and 26 nm the SPA observed is ten times larger. These results, which point to a lack of correlation between $M_S$ and SPA, are in agreement with by previous measurements [6] suggesting the existence of a narrow window of sizes for best absorption efficiency.

IV. DISCUSSION

Theoretical estimations indicate that the critical size for the single- to multi-domain transition is within the 30-40 nm for $Fe_3O_4$ particles.[7] The presence of two maxima observed in

χ''(T) curves agrees with the existence of two separate SPM unblocking process of the 17 nm and 26 nm particles, respectively), and indicates that both fractions are single-domain structure. Moreover, the thermally activated model of SPM particles could be used to obtain the activation energies $E_a = K_{eff}V$ for both 17 nm and 26 nm particles, yielding the same $K_{eff}$ value of 0.7(2) kJ/m$^3$. This result is in agreement with the fact of having two sets of particles with the same composition and same degree of interparticle interactions (i.e., coexisting in the same sample). From Table I it can be noted that the value of $K_{eff}$ together with the rest of the particles from GEYY (re-grown) samples are *lower* than the expected $K_1$ = 1.8 kJ/m$^3$ of bulk magnetite. Different sources of anisotropy (shape anisotropy, crystal defects, stress anisotropy, etc.) add to the total $K_{eff}$ values, so that $K_1$ is the lowest anisotropy value a given material can display, contrary to what is observed in Table I. However, morphological considerations from TEM images indicate that the symmetry of the particles should be considered since for d| 10 nm the particles display a clear cubic symmetry. Gittleman *et al.* have shown [8] that for cubic anisotropy and $K_1 < 0$ (as is the case of Fe$_3$O$_4$) the effective anisotropy is given by

$$K_{eff} = \frac{1}{12}K_1, \quad (1)$$

which yields values of magnetocrystalline anisotropy of $K_1$= 8.4 kJ/m$^3$, larger than the bulk value. The difference with the $K_1$ value can be explained by considering that these rough estimations do not include neither the additional contribution from dipolar interactions that are likely to exist in a powdered sample, nor the effects of disorder at the surface.

The heating efficiency of a magnetic colloid is measured by the heating power P of a given mass $m_{NP}$ of the constituent nanoparticles diluted in a mass $m_{LIQ}$ of liquid carrier, through the Specific Power Absorption

$$\Pi = \frac{P}{m_{NP}} = \frac{m_{LIQ}c_{LIQ} + m_{NP}c_{NP}}{m_{NP}}\left(\frac{\Delta T}{\Delta t}\right), \quad (2)$$

where $c_{LIQ}$ and $c_{NP}$ are the specific heat capacities of the liquid carrier. Since the concentrations of MNPs are usually in the range of 1 % wt., we can approximate [9] $m_{LIQ} c_{LIQ} + m_{NP} c_{NP} \approx m_{LIQ} c_{LIQ}$ and (2) can be written as

$$\Pi = \frac{c_{LIQ}\delta_{LIQ}}{\phi}\left(\frac{\Delta T}{\Delta t}\right), \quad (3)$$

with $\delta_{LIQ}$ and $\phi$ are the density of the liquid and the weight concentration of the MNPs in the colloid, respectively.

The use of different experimental conditions is usually the most restrictive aspect for comparing SPA data from different works.[10] The origin of (2) and (3) is clearly related to power conversion and heat exchange between an absorbing element (MNPs) and the liquid carrier of the colloid. Therefore, to obtain the actual SPA values for a given material, an *adiabatic* experiment is needed, where all the energy absorbed will be detected as a temperature increase due to the amount of heat generated within the sample. Since heating of a macroscopic amount of sample takes 10$^2$ to 10$^3$ seconds, high insulation is needed in order to grant adiabatic conditions during the whole experiment. The above considerations show that conclusions about the SPA values from *in vitro* or *in vivo* experiments (usually obtained from *isothermal* conditions) should be analyzed with care.

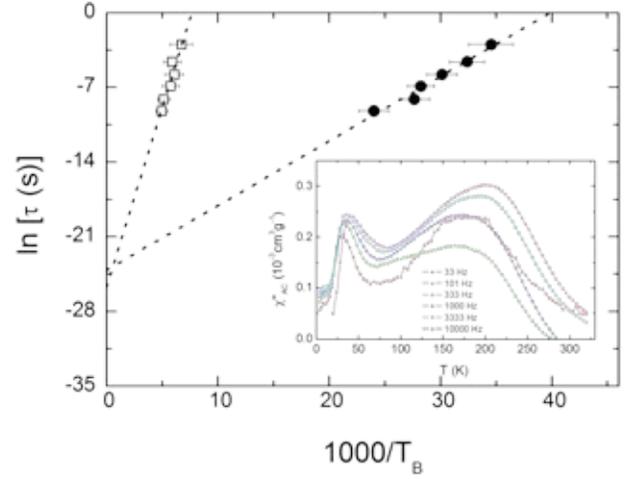

Fig. 2. Arrhenius plots from the magnetic ac susceptibility for sample GE03 with bi-modal distribution. The activation energies Ea for the two populations are different since Ea ~ V, but the same anisotropy constant $K_{eff}$ = 0.7 kJ/m$^3$ were obtained. Inset: χ'' curves at some frequencies showing the two-peak structure. The $K_{eff}$ values of all samples are shown in Table I.

The present experiments carried at 1% wt. concentration gave us a maximum SPA value of 130 W/g for GE03 sample, whereas for smaller and larger values the absorption was considerable lower. The strong dependence of SPA with particle size suggests that in addition to internal structure of the material, tailored synthesis routes will be necessary to tune the particle size to a given experimental condition. Although the origin of the large value in GE03 sample cannot be settled due to the coexistence of two definite populations, the same argument indicates that SPA could be further improved by separating both particle types. Since the 15 nm size particles

TABLE I
MAGNETIC PROPERTIES OF FE$_3$O$_4$ SAMPLES

| Sample | $<d>_{TEM}$ (nm) | $M_S$ † (emu/g) | SPA (W/g) | $K_{eff}$ § (kJ/m$^3$) |
|---|---|---|---|---|
| AP01 | 3.2 | 5.0 | 0 | 12.0 |
| AP02 | 7.3 | 11.0 | 5.2 | 2.3 |
| AP03 | 9.5 | 33.1 | 9.5 | 0.7 |
| GE01 | 11.5 | 36.4 | 5.0 | 0.9 |
| GE02 | 15.0 | 39.3 | 23.0 | 0.8 |
| GE03‡ | 17/26 | 41.5 | 130 | 0.7/0.7 |

Structural and magnetic data on the sample series with increasing particle size. For SPA measurements, all samples were standardized to a NP concentration of 1 % wt.
† Values at T = 300 K;
‡Bimodal size distribution, see text.
§ Obtained from the maxima of χ''(T) in the ac susceptibility data.

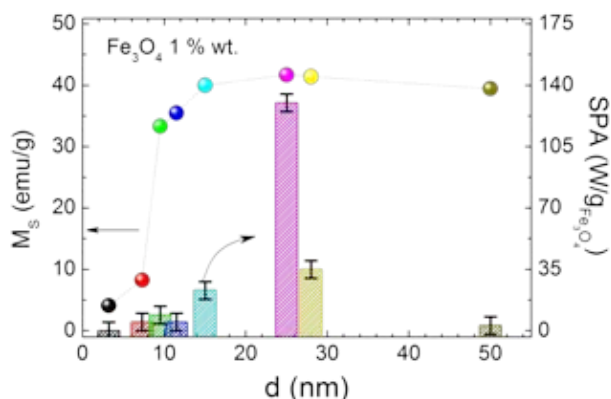

Fig. 3. Saturation magnetization values, MS, at room temperature (left axis) and the corresponding Specific Absorption Rate values (right axis) for magnetite particles with different mean particle diameters, d. It can be observed a jump in MS values for the larger particles, in a smaller d range than the observed increase of SPA.

of sample GE02 displayed a lower SPA, it is plausible that the origin of the SPA in GE03 could be originated mainly from the 26nm-particle system. Based on this idea and fitting the two size distributions from TEM histograms, we made a rough estimation of the contribution of a (hypothetically) pure 26 nm-size particle system, weighted for its mass contribution ($\approx$ 78% of the total mass), and found a SPA = 165 W/g. We mention here that the possibility of a large contribution from the (small) fraction of the largest particles could not be discarded (see the TEM histogram in fig. 1, where a noticeable number of d $\approx$ 30 nm particles could be counted). This range contains those particles with multi-domain structure, and if this were the case the mechanism should be related to domain wall dissipation within the particles, rather the coherent Néel relaxation observed in SPM particles. To clarify this point we have included in Figure 3 $M_S$ and SPA values for two samples with larger average particle diameters, up to 50 nm, obtained by coprecipitation. It can be observed that for larger sizes the SPA values are low, while Ms remain essentially constant. Although detailed work on larger particles is currently underway to systematically verify this hypothesis, current data support the landscape of Neel relaxation as the origin of SPA.

Although the SPA expression is already normalized for particle mass through the $\phi$ parameter, when $\Pi$ is measured against NP concentration the resulting values (not shown) were not constant as expected from (1) and (2), indicating that other effects, like increasing dipolar interactions among particles for concentrated colloids, should be added. Also, (3) is not normalized respect to the amplitude and frequency of the ac field applied in each experiment, and this is the main difficulty when comparing results from different groups in this field. The above considerations clearly show that many important parameters are not included in the phenomenological expressions (2) and (3), and improved models are needed for generating consistent data on SPA phenomena.

To summarize our results, we point that our study of SPA as a function of particle size shows that the average size and size distribution of the nanoparticles constituting a heating agent are central parameters for the design of efficient heating nanoparticles. Our data showed the absence of correlation between $M_S$ (or magnetic moment of a single particle) and SPA values for MNPs in the superparamagnetic regime. The data also suggest that the optimum particle diameter should be near the critical size for the single- to multi-domain transition for $Fe_3O_4$ phase, although the relation between SPA mechanisms and incipient domain walls is still to be determined. Detailed work on these issues is currently underway.

ACKNOWLEDGMENT

We are indebted to A. Labarta for fruitful discussions, and to P. Kyohara for the TEM images. This work was supported partially from Diputacion General de Aragon and Ministry of Education, Spain. Partial support from the Brazilian agency FAPESP is also acknowledged.